\documentclass[10pt,letterpaper,twocolumn,english,aps,showpacs]{revtex4}
\usepackage[T1]{fontenc}
\usepackage[latin9]{inputenc}
\usepackage{amstext}
\usepackage{amssymb}
\usepackage{graphicx}

\makeatletter

\pdfpageheight\paperheight
\pdfpagewidth\paperwidth

\newcommand{\lyxdot}{.}

\@ifundefined{textcolor}{}
{%
 \definecolor{BLACK}{gray}{0}
 \definecolor{WHITE}{gray}{1}
 \definecolor{RED}{rgb}{1,0,0}
 \definecolor{GREEN}{rgb}{0,1,0}
 \definecolor{BLUE}{rgb}{0,0,1}
 \definecolor{CYAN}{cmyk}{1,0,0,0}
 \definecolor{MAGENTA}{cmyk}{0,1,0,0}
 \definecolor{YELLOW}{cmyk}{0,0,1,0}
 }

\makeatletter
\gdef\@ptsize{0}
\let\@currsize\normalsize
\makeatother
\usepackage{setspace}
\makeatother

\usepackage{babel}
\begin{document}

\title{Global Phase Space of Coherence and Entanglement in a Double-Well BEC}

\author{Holger Hennig$^{1}$, Dirk Witthaut$^{2}$, David K.~Campbell$^{3}$ }

\affiliation{$^{1}$Department of Physics, Harvard University, Cambridge, MA 02138, USA}
\affiliation{$^{2}$Max Planck Institute for Dynamics and Self-Organization, 37073 G\"ottingen, Germany}
\affiliation{$^{3}$Department of Physics, Boston University, Boston, MA 02215, USA}

\date{\today}
\begin{abstract}
Ultracold atoms provide an ideal system for the realization of quantum 
technologies, but also for the study of fundamental physical questions
such as the emergence of decoherence and classicality in quantum
many-body systems. 
Here, we study the global structure of the quantum dynamics of bosonic 
atoms in a double-well trap and analyze the conditions for the generation
of many-particle entanglement and spin squeezing which have important
applications in quantum metrology.
We show how the quantum dynamics is determined by the 
phase space structure of the associated mean-field system and where 
true quantum features arise beyond this  `classical' approximation.
\end{abstract}


\maketitle
Maintaining and controlling the coherence of quantum systems over time
is one of the major challenges in contemporary physics. Low-temperature 
quantum gases trapped in optical lattices are an important instance of 
this challenge, for they provide versatile testbeds both for idealized models 
of exotic solid state systems and for applications in quantum optics and 
quantum information processing 
\cite{Bloch:2005uv,Bakr:2009bx,Bakr:2010gd,Simon:2011ep}.
A Bose-Einstein condensate (BEC) loaded in two linearly coupled wells, 
called a `BEC dimer' or `Bosonic Josephson Junction'  \cite{Javanainen:1986um}
is a particularly appealing case, as its quantum behavior is both theoretically 
tractable \cite{Micheli:2003bg,Polkovnikov:2003eh,Mahmud:2005jh,Witthaut:2008un} and experimentally accessible. In particular, BEC dimers have been shown to allow coherent manipulation of quantum states even on an atom chip,
enabling matter-wave interferometry \cite{Schumm:2005vn}, long phase coherence 
times and number squeezing \cite{Jo:2007jq,Esteve:2008vr,Gross:2010jn}, and 
proposed chip-based gravity detectors \cite{Hall:2007bh}.
Related questions have been addressed in the context of the mathematically equivalent Lipkin-Meshkov-Glick model \cite{Orus:2008kr}.

For large atom numbers, the coarse dynamics of the BEC dimer is 
excellently described by a `classical' mean-field approximation.
Recent experiments have precisely mapped out the emerging 
classical phase space structure \cite{Albiez:2005wk,Zibold:2010el}, 
showing macroscopic quantum oscillations as well as the emergence 
of self-trapped states through a `classical' bifurcation. 
Recent theoretical studies have focused on strategies to prepare highly 
entangled Einstein-Podolsky-Rosen (EPR) states in the BEC dimer
\cite{Polkovnikov:2002ew,Hillery:2006je,He:2011jf,BarGill:2011hd}.
These truly quantum phenomena appear to depend on the structure of the associated classical phase space  \cite{Polkovnikov:2002ew,Micheli:2003bg,Polkovnikov:2003eh,Mahmud:2005jh,Witthaut:2008un}, but what is the exact correspondence?

Here, we introduce a {\em global} phase space (GPS) picture of the dynamics of the quantum dimer, focussing on entanglement, spin squeezing and decoherence of the atoms. 
Our global picture shows that both the time evolution of entanglement and coherence depend very fundamentally on the initial coherent state. In particular, we find that self-trapping (ST) excellently supports many-particle entanglement.
 To link quantum and classical aspects of the dynamics, we approach the problem from three perspectives.
 First, for the full quantum problem, we employ the two-site Bose-Hubbard model
and solve for all of the above quantum observables as functions of time. Second, 
we recall that the standard mean-field treatment of the dimer -- the two-site 
Gross-Pitaevski equation -- corresponds to an integrable classical dynamical 
system, and we discuss the nature of the trajectories in this system. Third, we use a recently developed semiclassical `Liouville dynamics approach' (LiDA) \cite{Trimborn:2009up}, 
which we argue is intermediate between the fully quantum and the classical systems. 
Comparing these three perspectives, our GPS approach
provides an intuitive understanding of the quantum observables mentioned above and 
insights into true quantum effects beyond the mean-field or semiclassical approaches.

\emph{Quantum and semiclassical dynamics.--}
The coherent dynamics of the Bose-Einstein condensate dimer
is given by the two-mode Bose-Hubbard Hamiltonian (BHH)
\begin{equation}
  H = - J (\hat a_1^\dagger \hat a_2 + \hat a_2^\dagger \hat a_1) 
   + \frac{U}{2} \left( \hat a_1^{\dagger 2} \hat a_1^2 +  \hat a_2^{\dagger 2} \hat a_2^2 \right)
  \label{eq:BHH}
\end{equation}
where $\hat a_j$ denotes the annihilation of one bosonic atom in state
$j$. The modes can be either realized by two sites in a double-well trap
\cite{Gati:2006ws,Gati:2006cl}) or two internal states of the atoms 
\cite{Zibold:2010el}.
The tunneling rate $J$ and the on-site interaction $U$ can be tuned 
individually, e.g., via a Feshbach resonance or by changing the depth
of atom trap \cite{Inouye:1998vx,Bloch:2005uv,Syassen:2008um,Gross:2010jn,Zibold:2010el,Nicklas:2011el}. Throughout this paper we set $\hbar = 1$, thus measuring
all energies in frequency units. If not stated otherwise, we assume that
$J = 10 \, {\rm s}^{-1}$, which is a common setting in ongoing experiments
\cite{Gross:2010jn,Zibold:2010el,Nicklas:2011el}, and we assume a total
number of $N=40$ atoms.

\begin{figure*}

\includegraphics[width=0.62\columnwidth]{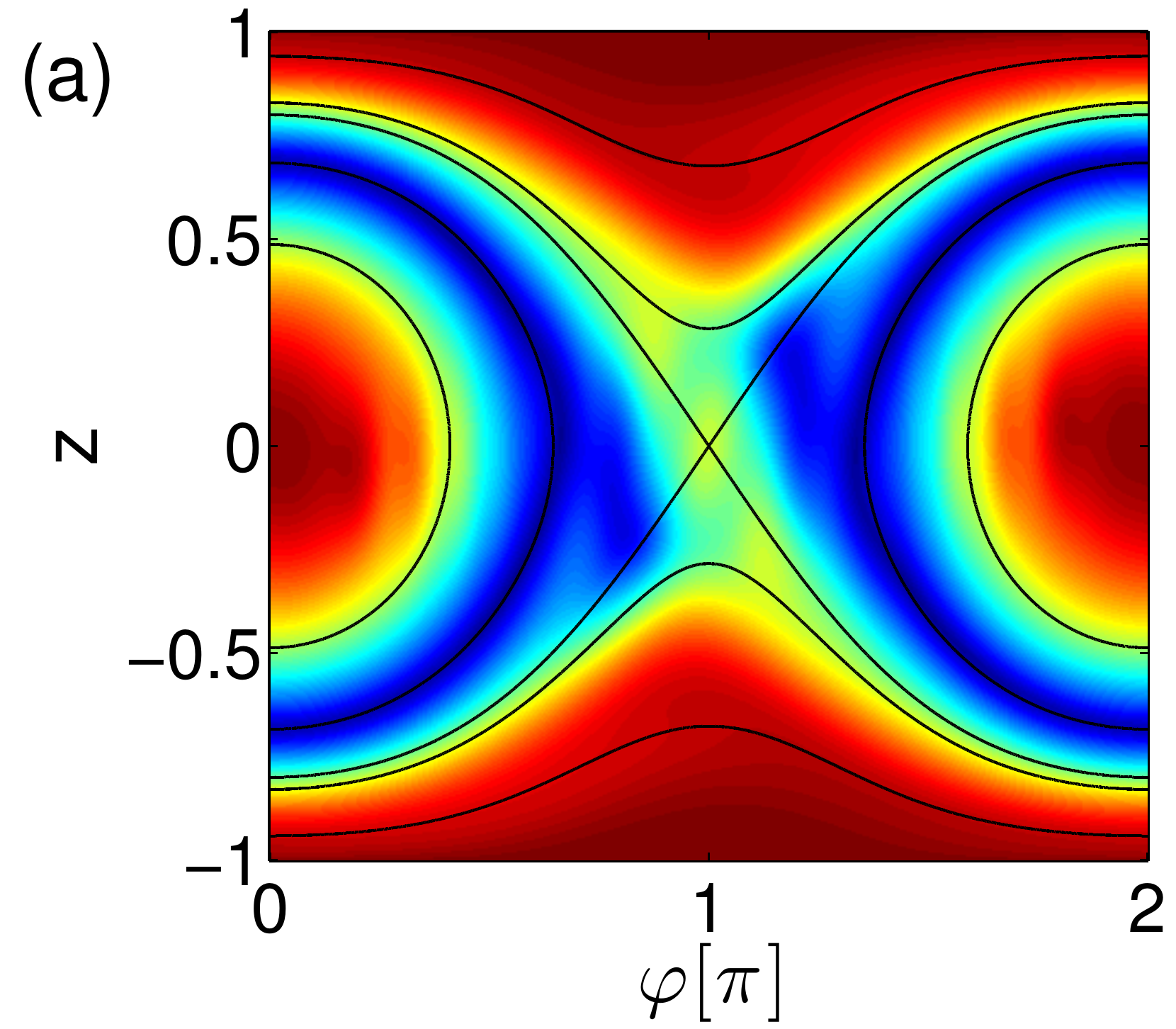}
\includegraphics[width=0.62\columnwidth]{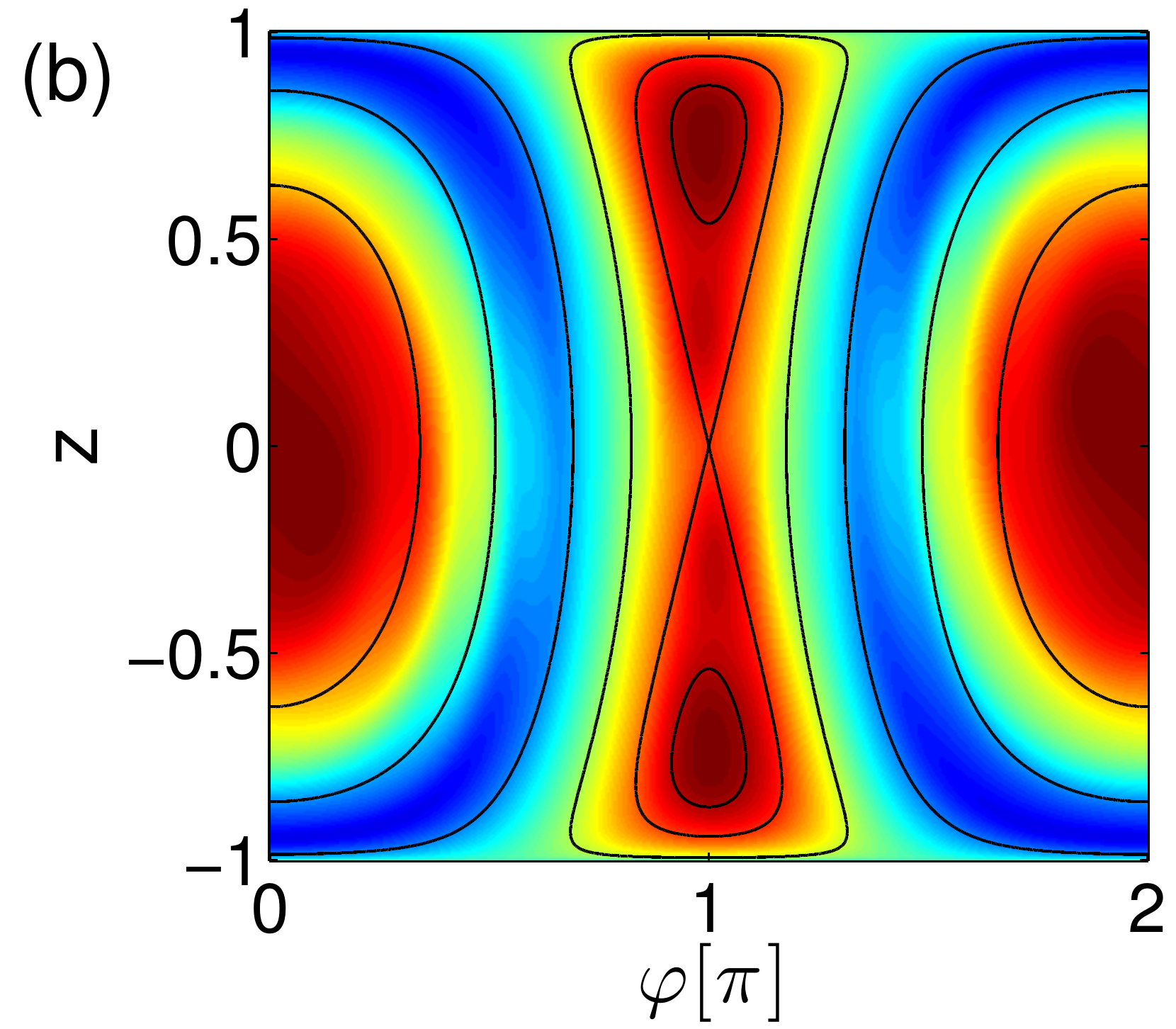}
\includegraphics[width=0.62\columnwidth]{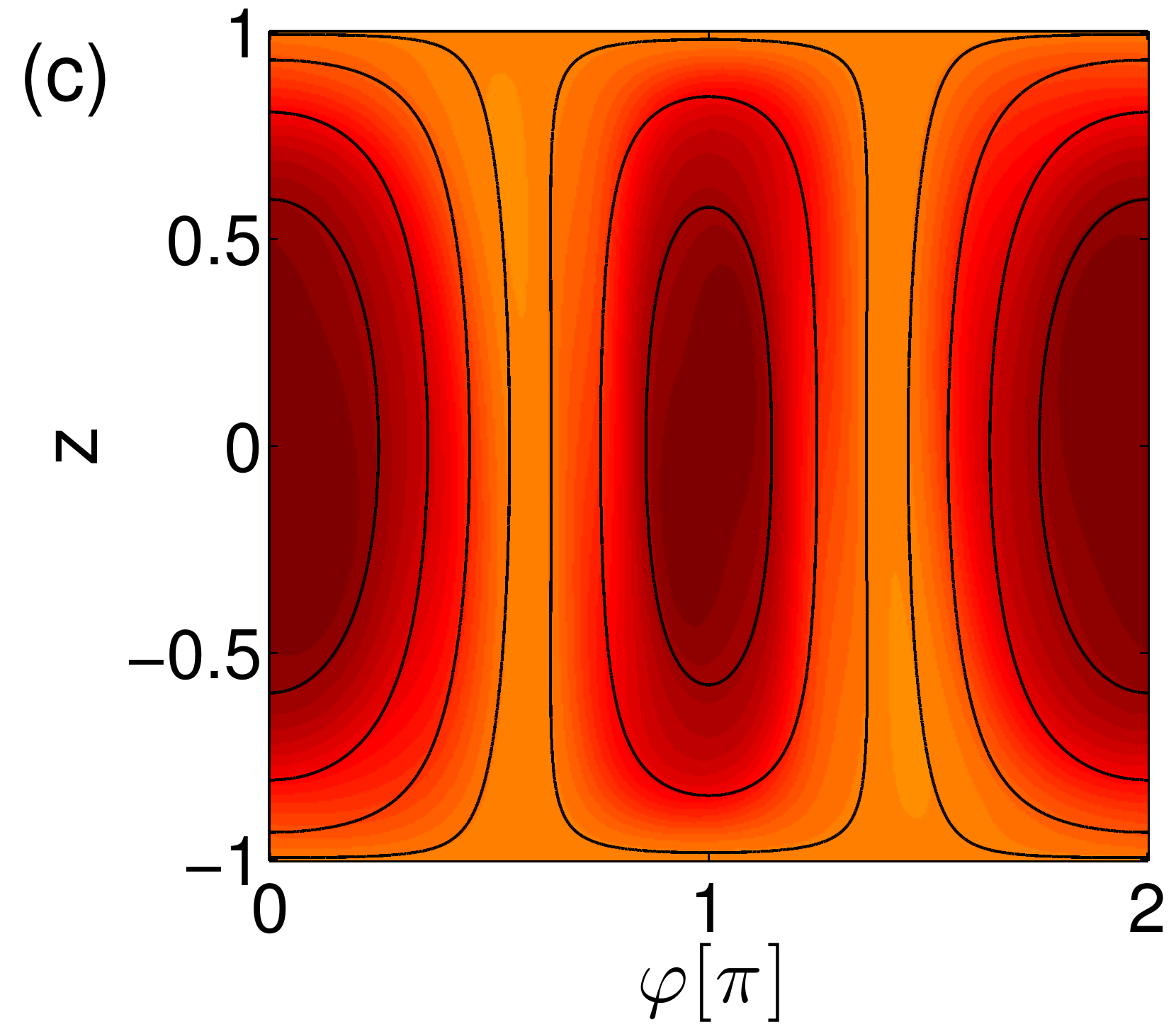}
\includegraphics[width=0.098\columnwidth]{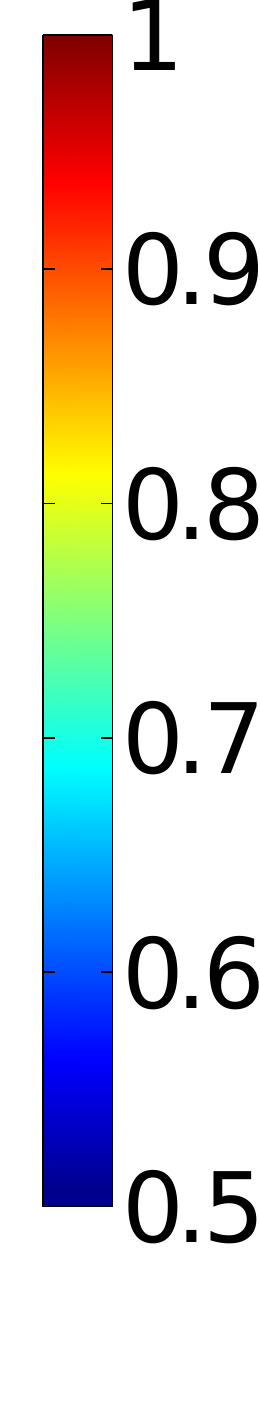}

\caption{
Global phase space structure of the Bose-Hubbard dimer for \textbf{(a)}
$\Lambda=5$, \textbf{(b)} $\Lambda=1.5$,  \textbf{(c)} $\Lambda=0.5$.
The color code shows the condensate fraction $c_{\tau}(\varphi,z)$ at
$\tau\!=\!1\,$s as a function of the initial state $|\varphi,z\rangle$.
Lines of equal condensate fraction mimic the classical trajectories 
(black lines), despite the clear qualitative differences for the three
values of $\Lambda$ shown.
However, we observe strong deviations in the coherence properties of the 
quantum state, such as S-symmetry breaking and enhanced condensate 
fraction near the unstable fixed point $F_2=(\pi,0)$, which are clearly beyond
the classical mean-field description. 
\label{fig:qps}}
\end{figure*}

The mean-field counterpart of the Bose-Hubbard dimer is a two-mode 
Gross-Pitaevskii or nonlinear Schr\"odinger equation. This equation can
be rewritten as an integrable classical Hamiltonian system with the 
Hamiltonian \cite{Raghavan:1999ud}
\begin{equation}
  H_{cl}(z,\varphi)=\frac{\Lambda z^{2}}{2}-\sqrt{1-z^{2}} \, \cos(\varphi)\,.\label{eq:classicalHamiltonian}
\end{equation}
The mean-field (classical) phase space consists of the two conjugate variables, 
which following previous usage \cite{Raghavan:1999ud}, we have chosen to be 
the relative phase $\varphi=\varphi_{1}-\varphi_{2} \in [0,2\pi)$ between 
the two wells and the population imbalance $z=(N_{1}-N_{2})/N \in [0,1]$,
where $N_{1}$ and $N_{2}$ denote the number of atoms in each well and 
$N=N_{1}+N_{2}$ is the total atomic population. The classical trajectories 
simply follow the lines of constant (conserved) energy $H_{cl}(z,\varphi) = const.$
They are determined by the initial position $(\varphi_{0},z_{0})$ in phase 
space and the ratio of the interaction and the tunneling energy $\Lambda=UN/(2J)$.

The Gross-Pitaevskii equation implicitly assumes a pure BEC
at all times. Some quantum features beyond this rough 
approximation, in particular the quantum mechanical spreading over time,
can be included if a quantum state $|\Psi\rangle$ is represented by a 
(quantum) phase-space density such as the Husimi function
$Q(\varphi,z)=|\langle\varphi,z|\varphi_{0},z_{0}\rangle|^{2}$
instead of a single trajectory. 
Here, $|\varphi ,z \rangle$ denotes an \emph{atomic coherent state} 
\cite{Zhang:1990fy}, which is nothing but a pure BEC.
As shown in \cite{Trimborn:2009up}, the dynamics of the Husimi function follows 
a classical Liouville equation with the Hamiltonian (\ref{eq:classicalHamiltonian}) 
plus quantum correction terms vanishing as $1/N$.
In the semiclassical LiDA \cite{Trimborn:2009up} we thus represent an
initial state by an {\it ensemble} of trajectories, whose initial positions
are distributed according to the Husimi function of the initial quantum state.

\emph{Global phase space structure.--}
We shall analyze the \emph{global} phase space structure of the
Bose-Hubbard dimer with special respect to its classical and quantum properties.
Therefore, we consider the dynamics of an initially pure BEC as a function 
of the parameters $z_{0}$ and $\varphi_{0}$ with a focus on the
purity of the condensate and the emergence of entanglement and spin-squeezing.
The purity is measured by the condensate fraction $c_{\tau}$ defined as 
the maximum eigenvalue of the reduced single-particle density matrix 
(SPDM) $\rho_{ij}=\langle a_{i}^{\dagger}a_{j}\rangle/N$ 
\cite{PETHICK:2008tn,Witthaut:2008un,Trimborn:2009up}. 
A related GPS approach was introduced in terms
of the phase space entropy \cite{Mirbach:1995ul}, which provides great insight into global dynamical properties of the system.

The global phase space structure of the Bose-Hubbard dimer is shown in
Fig.~\ref{fig:qps}, where the condensate fraction $c_{\tau}(\varphi,z)$ at 
time $\tau\!=\!1\,$s is plotted as a function of the initial state
$|\varphi_0,z_0\rangle$ for three different values of $\Lambda$.
The corresponding classical mean-field dynamics is overlaid as solid 
black lines. For $\Lambda < 1$, the atoms show simple Rabi oscillations 
between the wells with stable fixed points at $F_1=(0,0)$ and $F_2=(\pi,0)$. 
According to the average phase $\bar{\varphi}$, the oscillations are referred 
to as `zero-phase' or plasma oscillations ($\bar{\varphi}=0$) around $F_1$ and  
`$\pi$-phase' oscillations ($\bar{\varphi}=\pi$) around $F_2$.
The classical dynamics undergoes a bifurcation at $\Lambda=1$, separating 
the Rabi ($0\!<\!\Lambda\!<\!1$) and Josephson ($\Lambda\!>\!1$) regimes. 
The fixed point $F_2$ becomes hyperbolically unstable for  $\Lambda \!>\! 1$, 
bifurcating into two self-trapping (ST) fixed points at 
$F_\textit{ST}=(\pi,z_\textit{ST})$, where 
$z_\textit{ST}=\pm \sqrt{1-1/\Lambda^2}$ \cite{Raghavan:1999ud}.
In addition to  $\pi$-phase  ST with average phase $\pi$ (Fig.~\ref{fig:qps}b), also `running phase' ST appears for $\Lambda>2$ (see Fig.~\ref{fig:qps}a).

Comparing the condensate fraction $c_\tau(\varphi,z)$ with the mean-field 
dynamics we find that lines of equal  $c_\tau(\varphi,z)$ mimic the classical 
trajectories in many respects.
The GPS mapping for $c$ clearly reflects  zero- and $\pi$-phase oscillations
and the stable fixed points (Fig.~\ref{fig:qps}(c)) as well as ST (Fig.~\ref{fig:qps}(a-b)).
A drastic drop in the condensate fraction is observed in particular near the classical 
separatrices, indicating a deviation from a pure BEC and thus the failure of a simple 
mean-field description.
Notably, this loss of coherence is not strongest 
at the unstable fixed point, cf.~Fig.~\ref{fig:qps}(a-b). 
Furthermore, the quantum dynamics breaks the
symmetry $S\!:\! (\varphi,z)\to (\varphi,-z)$, as
$c_{\tau}(S(\varphi,z))\ne c_{\tau}(\varphi,z)$. In the classical limit $S$ is preserved:
$H(S(\varphi,z)) = H(\varphi,z)$, see Eq.~(\ref{eq:classicalHamiltonian}).
Of course, both the classical and
the quantum dynamics are symmetric with respect to $(\varphi,z)\to(-\varphi,-z)$,
which translates to relabeling wells $1\leftrightarrow2$ without
changing the initial state.

\begin{figure}
\includegraphics[width=0.48\columnwidth]{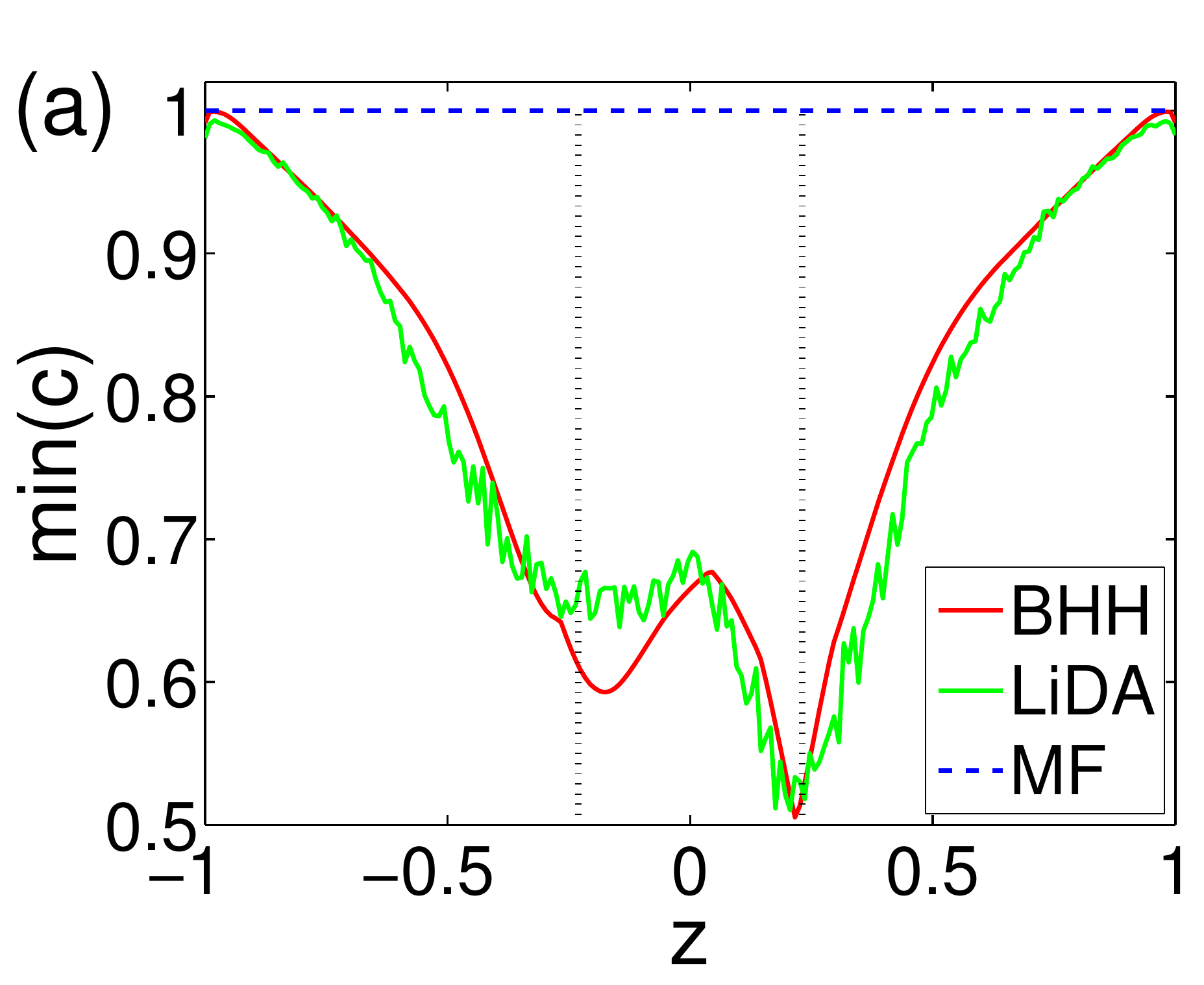}
\includegraphics[width=0.48\columnwidth]{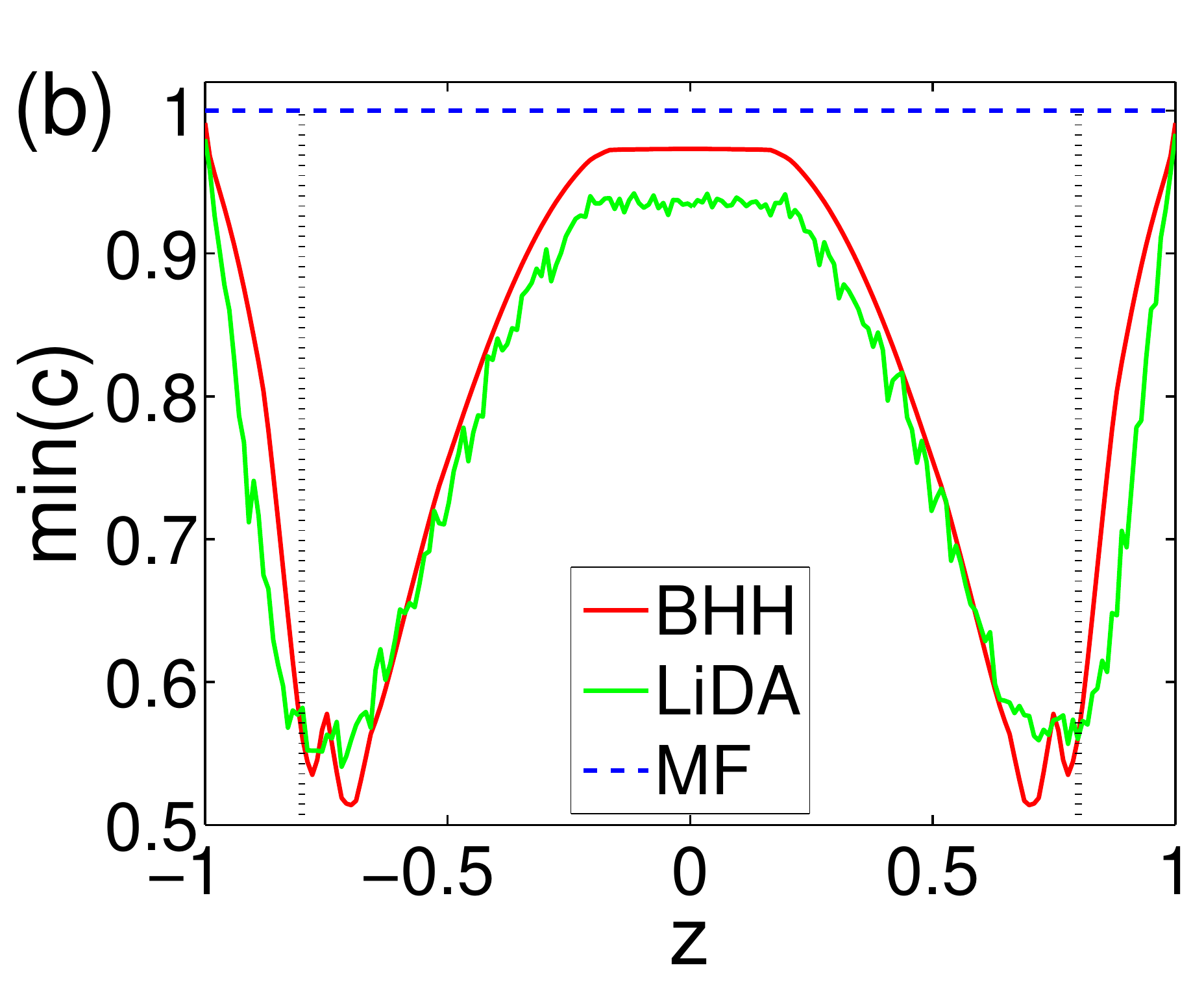}
\includegraphics[width=0.55\columnwidth]{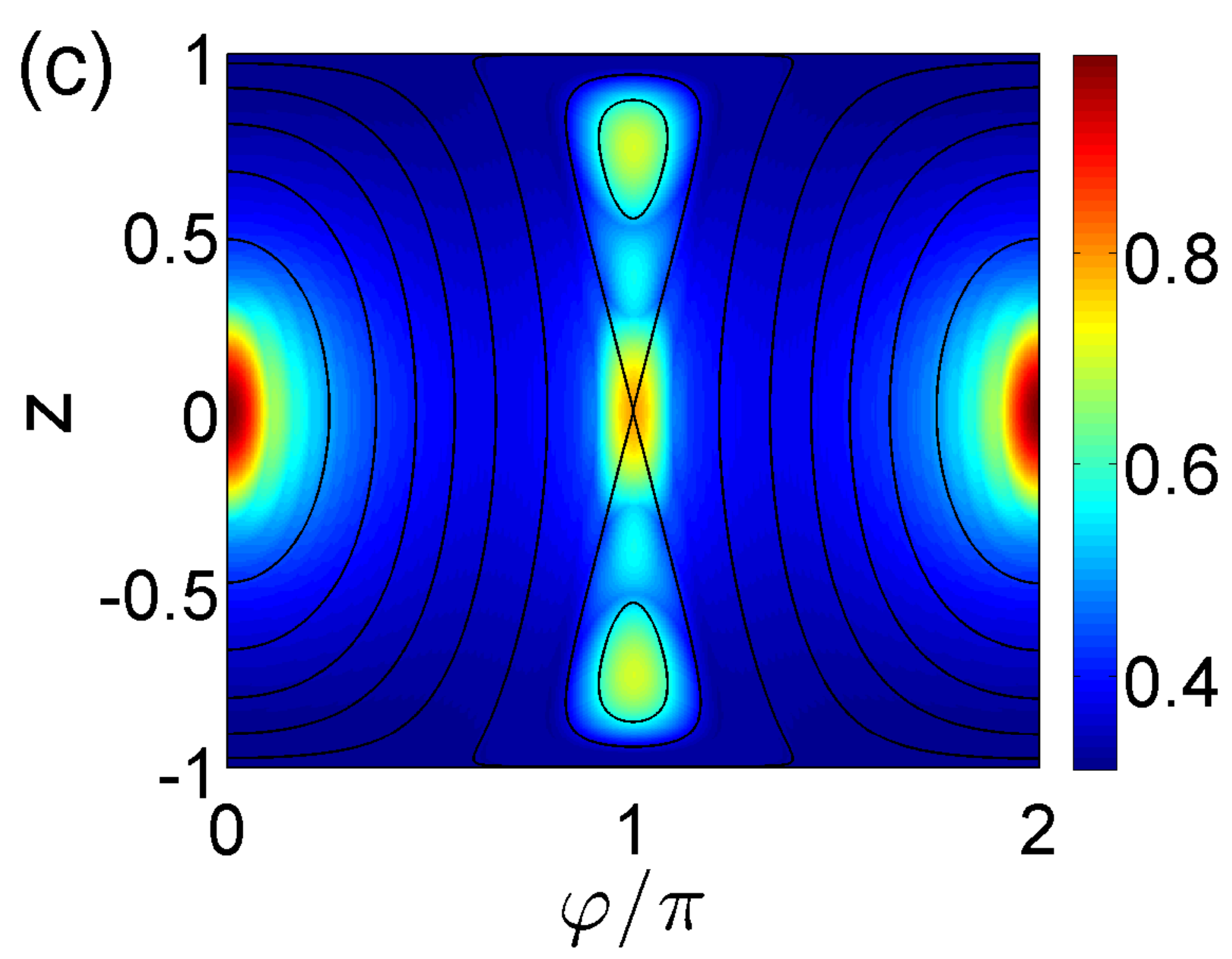}

\caption{The GPS approach reveals novel  features of quantum 
dynamical origin beyond the classical mean-field description and 
the semiclassical LiDA.
\textbf{(a-b)} Sections through the GPS 
for $\Lambda\!\!=\!\!5$ comparing quantum results (based upon the BHH, red curves) with LiDA (green curves). The mean-field (MF) description assumes a pure BEC (blue dashed line). 
\textbf{(a)} Shown is the condensate fraction $\min(c_\tau(0.85\pi,z))$ 
in the time interval $0 <\!\! \ \tau < 0.5\,$s. 
The point $(0.85\pi,0)$ is near the hyperbolic fixed point. The LiDA reproduces $S$-symmetry breaking, but can distinctly deviate from the quantum results near the hyperbolic fixed point and the separatrices (dashed vertical lines).
\textbf{(b)} 
For $\min(c_{\tau}(0,z))$ the LiDA deviates from the quantum dynamics for $|z|\lesssim 0.8$, i.e.~near the separatrices (which are located at $z=\pm0.8$, dashed lines) 
\textbf{(c)} Large condensate fraction $c_\tau$ in the vicinity of the classical fixed
points is reproduced qualitatively by a time-independent
measure. The color code shows the maximum overlap of the initial
state with a stationary state, $A(\varphi,z)=\max_{n}|\langle\varphi,z\,|\, E_{n}\rangle|$
for $\Lambda=1.5$.
However, several quantum features depicted in the GPS, such as the symmetry 
breaking, are not reproduced by this measure, indicating 
that these are of dynamical origin.
\label{fig:quantum_features}}
\end{figure}

The correspondence of quantum and classical phase space structure is
more than a qualitative coincidence.
In Fig.~\ref{fig:quantum_features} (a,b) we compare the quantum results (based upon the BHH) for the 
minimum condensate fraction with the prediction of the semiclassical LiDA. The good
agreement of both quantities reveals that the loss of quantum coherence 
 $c_\tau(\varphi,z)$ can be mostly attributed to the \emph{classical} spreading of
Husimi function. Around a stable fixed point neighboring trajectories remain
close for all times, such that there is no spreading.

But why does the quantum coherence remain reasonably high near the 
unstable fixed point? We recall that Eq.~(\ref{eq:classicalHamiltonian})
can be mapped to a pendulum with variable length $l(z)=\sqrt{1-z^{2}}$ 
and angular velocity $z=\dot{\varphi}$. Within that picture, the unstable 
fixed point $F_2=(\pi,0)$ is reached when the pendulum is in its upright 
position with no angular velocity. 
Close to the unstable fixed point, the classical dynamics  
slows down asymptotically (``freezes'');
therefore spreading near the unstable fixed point is slow.
Moreover, when the trajectory is moving towards $F_{2}$ 
the dynamics slows down, while moving away from $F_{2}$ 
the pendulum accelerates, which breaks the $S$-symmetry.
On the contrary, close to 
the separatrix, the 
classical dynamics leads to delocalization in phase space which
induces decoherence of a many particle state 
\cite{Mirbach:1995ul,Trimborn:2009up}.

To study to what extent these features can be captured in a time-independent
framework, we analyze how much a state $|\varphi,z\rangle$ overlaps with an eigenstate $|E_{n}\rangle$ of the 
Bose-Hubbard dimer.
Figure \ref{fig:quantum_features} (c) shows the maximum overlap 
$A_{\text{max}}(\varphi,z)=\max_{n}|\langle\varphi,z\,|\, E_{n}\rangle|$
for the same parameters as in Fig.~\ref{fig:qps} (b). The overlap with
the eigenfunctions is high at the stable fixed points and minimal along 
most parts along the separatrix. This reveals one possible mechanism to 
guarantee long-time coherence of a BEC: If the overlap approaches unity, 
the initial state is almost stationary such that the condensate fraction remains 
virtually constant.
This is consistent with (but the converse of) the results reported in
refs.~\cite{Mirbach:1995ul,Trimborn:2009up}, where it was shown
that delocalization in phase space induces decoherence of a
quantum state.
Moreover, there is as well an eigenfunction localized around the unstable 
fixed point $F_{2}$, which explains the surprisingly slow decoherence of 
the quantum state localized at this point.

\begin{figure}
\includegraphics[width=0.51\columnwidth]{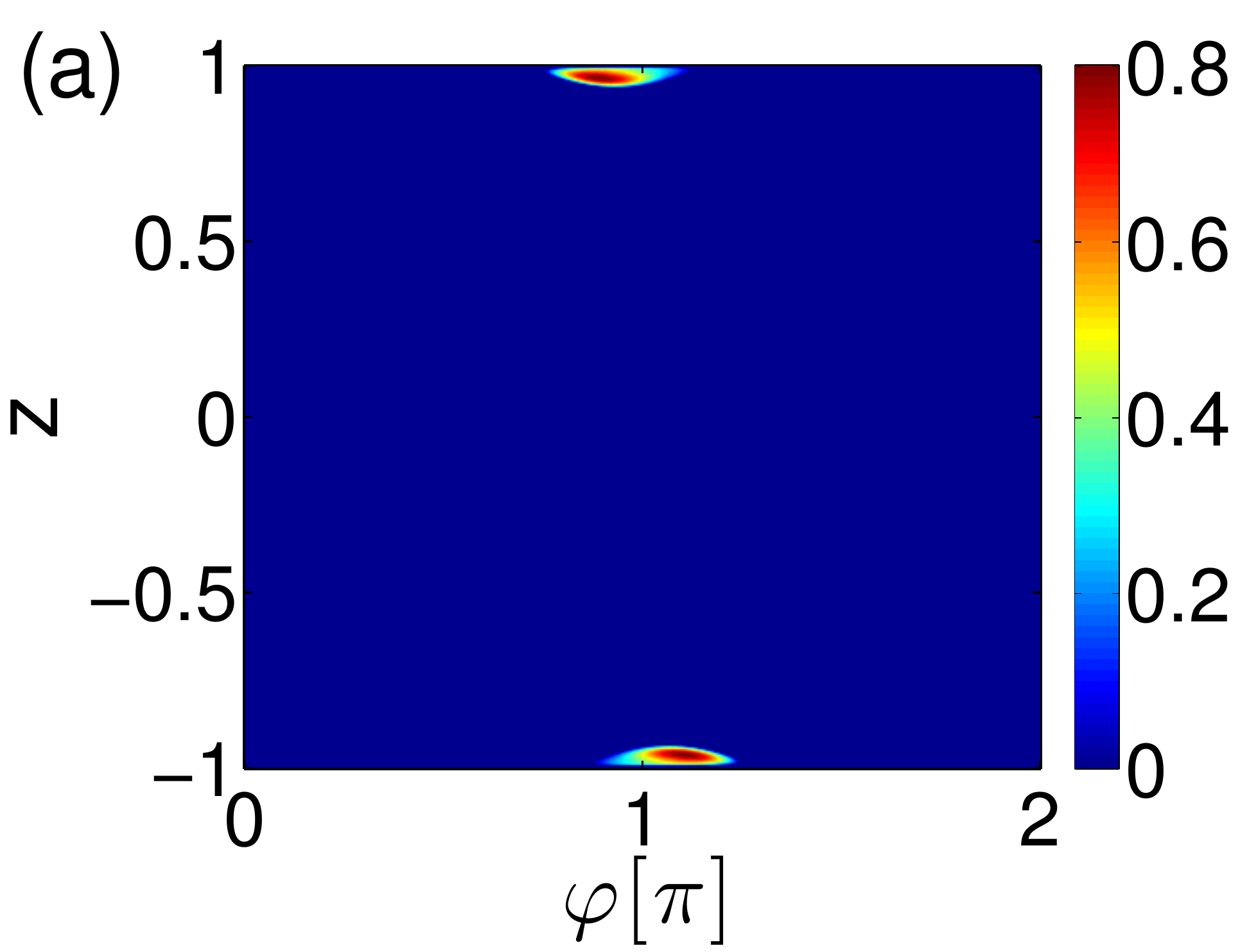}
\includegraphics[width=0.49\columnwidth]{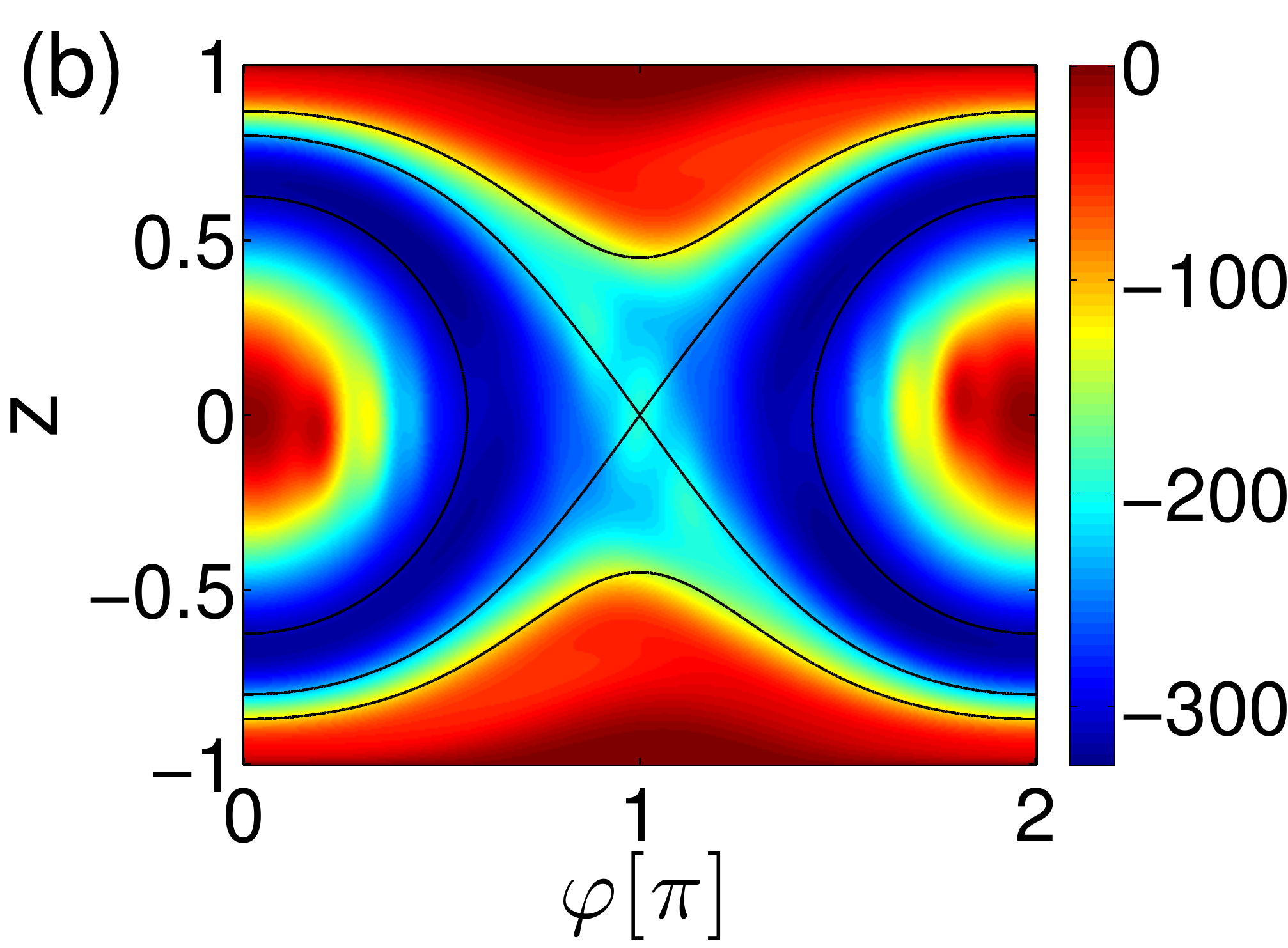}
\includegraphics[width=0.49\columnwidth]{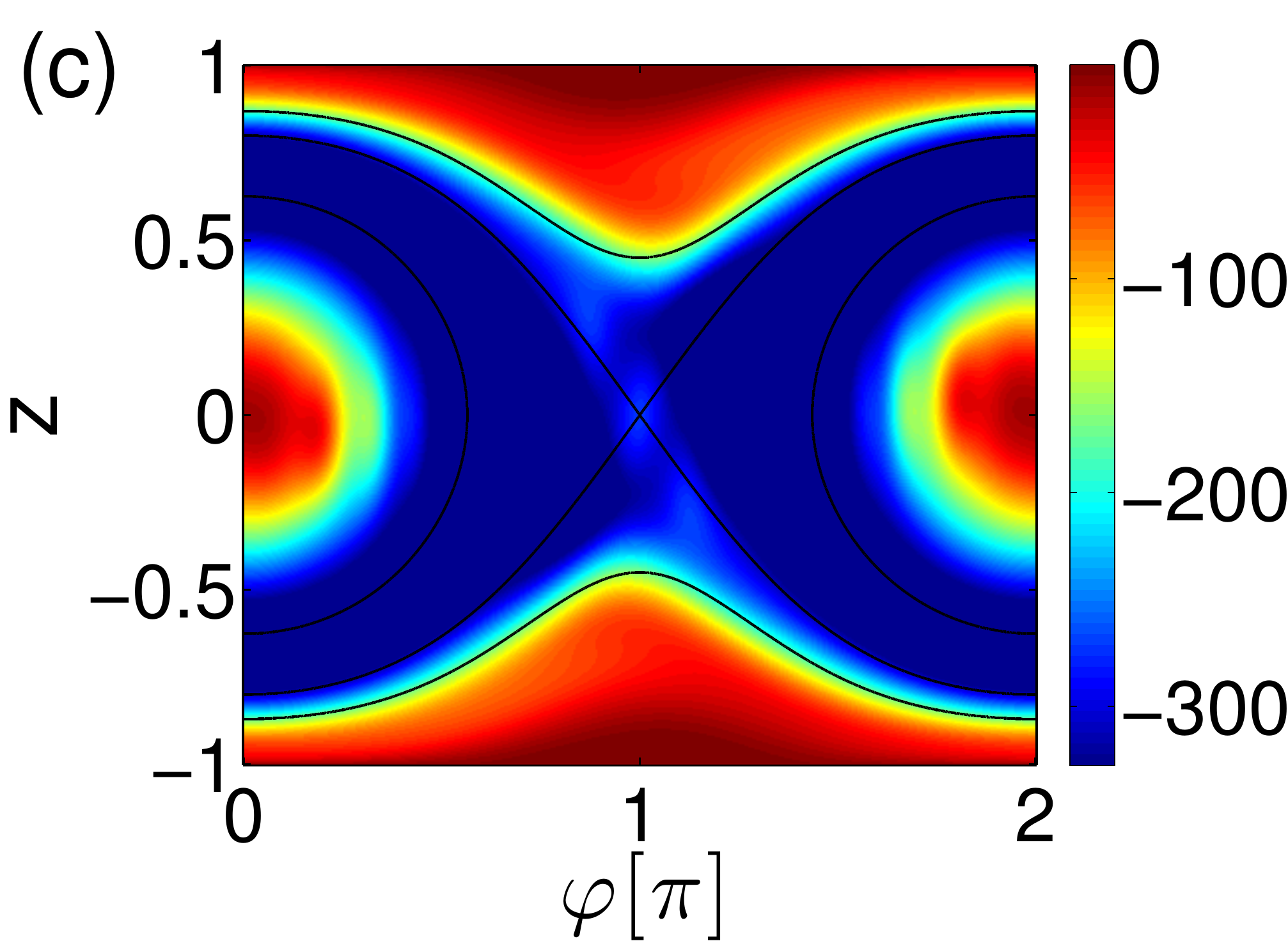}
\includegraphics[width=0.49\columnwidth]{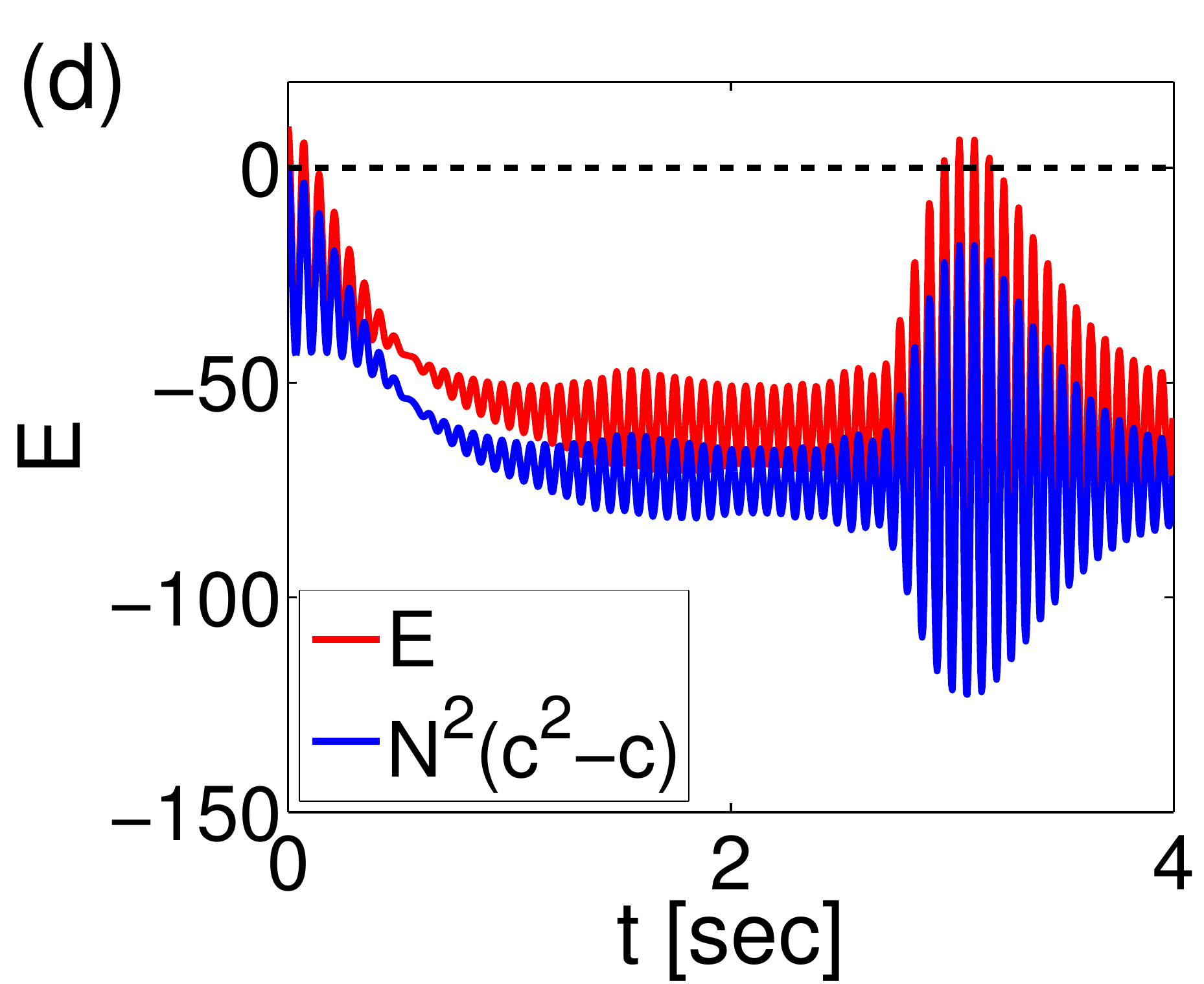}
\includegraphics[width=0.49\columnwidth]{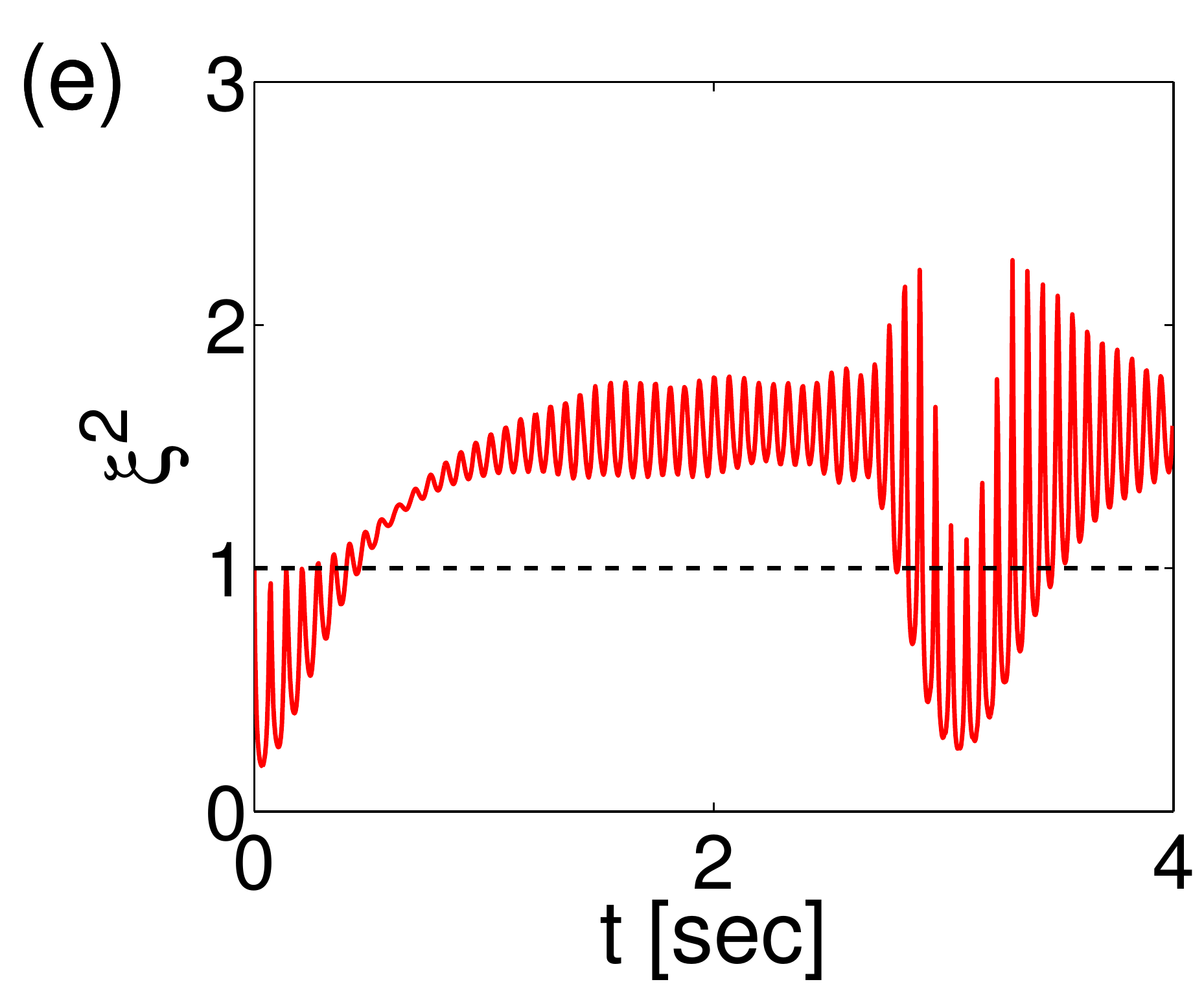}

\caption{
Global phase space structure of EPR entanglement and spin squeezing 
for $\Lambda=5$. 
\textbf{(a)} EPR entanglement $E>0$ is found at $\tau=1\,$s solely around the ST fixed points.
\textbf{(b,c)} The entanglement parameter $E_\tau(\varphi,z)$ \textbf{(b)} and the 
semiclassical approximation $E_{sc}=N^2(c^2-c)$ \textbf{(c)} at time 
$\tau=1\,$s are shown as a function of the initial state $|\varphi,z\rangle$. While these snapshots depict the GPS at $\tau=1\,$s, the movie in the supplementary material shows the time evolution of the GPS in timesteps of $0.025\,$s.
\textbf{(d,e)} Time evolution of EPR entanglement \textbf{(d)} and spin squeezing \textbf{(e)} 
for an initially pure BEC $|\varphi,z\rangle=|0,0.2\rangle$. The semiclassical 
approximation reproduces the overall behavior quite well but exhibits an almost constant shift. 
A revival of entanglement is observed at $t\approx3\,$s.  
Spin squeezing ($\xi^2<1$) is found to be in excellent agreement with the EPR entanglement 
result ($E>0$) at $t\approx3\,$s. 
 \label{fig:Spin-squeezing} }
\end{figure}

\emph{Entanglement.--}
A distinguishing feature of experiments with two-mode BECs is that the quantum 
state of the atoms can be manipulated with astonishing precision.
In particular, the atoms can be strongly entangled, which enables important
applications in precision quantum metrology 
\cite{Wineland:1994vq,Sorensen:2000vw,Gross:2010jn}.
For the important special case of EPR entanglement, a simple
criterion reads $E>0$, where 
$E = | \langle \hat a_1^\dagger \hat a_2 \rangle  |^2  -  
\langle \hat a_1^\dagger \hat a_1 \hat a_2^\dagger \hat a_2 \rangle$ 
\cite{Hillery:2006je,He:2011jf}.
Proceeding in the same way as for the condensate fraction, we obtain a 
GPS picture for the EPR entanglement shown in Fig.~\ref{fig:Spin-squeezing} 
(a,b). The measure $E_\tau(\varphi,z)$ again excellently mimics the 
classical phase space trajectories, however, an actual entangled state 
(i.e., $E>0$) is found at time $\tau=1\,$s only near the classical ST 
fixed points $(\pi,z_\textit{ST})$.
Movies showing the temporal evolution of $E_\tau(\varphi,z)$ for $\tau\in [0,3]\,$s can be found in the supplementary material.
Strikingly, the GPS for entanglement measured at different times $\tau$ 
unveils the following behavior: If entanglement appears
on timescales long compared to the period of the plasma oscillations, 
then entanglement is concentrated around one or more fixed points  in phase space, 
which, surprisingly, includes the unstable fixed point $F_2$.

Given the apparent similarity of the GPS for $E$ and the condensate fraction 
$c$, we ask, how is $E$ related to $c$? By approximating 
$\langle \hat a_1^\dagger \hat a_1 \hat a_2^\dagger \hat a_2 \rangle \approx 
\langle \hat a_1^\dagger \hat a_1 \rangle \langle \hat a_2^\dagger \hat a_2 \rangle$
we find a semiclassical measure 
$E_\text{sc}= | \langle \hat a_1^\dagger \hat a_2 \rangle  |^2  -  
\langle \hat a_1^\dagger \hat a_1 \rangle \langle \hat a_2^\dagger \hat a_2 \rangle 
= N^2 (c^2-c)$. In Fig.~\ref{fig:Spin-squeezing}(c) a GPS for 
$E_\text{sc}$ is shown for the same parameters, with remarkable 
qualitative agreement with the exact quantum results for $E$. 
In Fig.~\ref{fig:Spin-squeezing}(d) we report the temporal 
evolution of $E$ and $E_\text{sc}$ of the state $| 0.2,0 \rangle$ 
which  shows surprisingly good agreement except for an offset. However, 
as $E_\text{sc}\le0$ by definition, it does not serve as 
an entanglement criterion.

Spin squeezing indicates a form of many-particle entanglement which is particularly 
important for quantum metrology \cite{Wineland:1994vq,Gross:2010jn}.
A state is spin-squeezed if the quantum uncertainty in a Bloch sphere
representation is smaller than that of an atomic coherent state, i.e., a pure BEC.
This representation is defined via the operators
$\hat{J}_{x}  \!=\! \frac{1}{2}(\hat a_{2}^{\dagger} \hat a_{1}\! + 
\! \hat  a_{1}^{\dagger} \hat a_{2})$, 
$\hat{J}_{y}  \!=\!  \frac{i}{2}(\hat a_{2}^{\dagger} \hat a_{1}\! - 
\! \hat a_{1}^{\dagger} \hat  a_{2})$, and
$\hat{J}_{z}  \!=\!  \frac{1}{2}(\hat a_{1}^{\dagger} \hat a_{1}\! - 
\! \hat  a_{2}^{\dagger} \hat  a_{2})$,
which form an angular momentum algebra with quantum number $J=N/2$ 
\cite{Zhang:1990fy,Trimborn:2009up,PETHICK:2008tn}. 
A quantum state is spectroscopically squeezed if the parameter
\cite{Sorensen:2000vw}
\begin{equation}
  \xi^{2}:=\frac{N(\Delta \hat{J}_{n_{1}})^{2}}{\langle \hat{J}_{n_{2}}\rangle^{2}+\langle \hat{J}_{n_{3}}\rangle^{2}} 
  \label{eq:squeezing}
\end{equation}
is smaller than one. Here, $\hat{J}_{n_i}\!=\!n_i \cdot \hat{J}$ is the projection 
of the total angular momentum operator $\hat{J}$ onto $n_i$, where 
$n_{1,2,3}$ are mutually orthogonal unit vectors and $\Delta \hat{J}_{n}$ is 
the variance of $\hat{J}_n$. In Fig.~\ref{fig:Spin-squeezing} (e) we report 
the dynamical evolution of $\xi^2$ for an initially pure BEC 
$| 0.2,0 \rangle$ in excellent agreement with 
$E(t)$ (d), including the entanglement revival at $t\approx 3\,$s.

\emph{Discussion.--}
While chaotic classical dynamics has been shown typically to lead to a fast
(exponential) decay of the coherence of a many-particle state \cite{Castin:1997vx},
little is known about the coherence of quantum self-trapped states (especially
in a high dimensional phase space), where the classical counterparts
are (quasi-)periodic orbits, although some insight can be provided by 
semiclassical approaches beyond the mean-field limit 
\cite{Vardi:2001vw,Witthaut:2011kv,Trimborn:2011gy}. 
Fig.~\ref{fig:quantum_features} reveals one aspect of the coherence 
of self-trapped states. For $z$ near $\pm1$ the many-particle state is 
close to a pure condensate at time $\tau$, whereas the condensate 
fraction drops drastically well before the separatrix (dashed line) is 
reached. In Fig.~\ref{fig:qps}(a-b) the dark red regions around 
the fixed points $F_\text{ST}$ further confirm that the self-trapped 
regions maintain high condensate fraction. Hence, ST is a mechanism 
to preserve coherence of many-particle states. In contrast, in the vicinity 
of the separatrix the condensate fraction falls off sharply.

We have analyzed the connection between quantum observables 
of the BEC dimer and the structure of the underlying classical 
phase space, including fixed points, separatrices, and  ST. The 
question remains: how well do our results carry over beyond the 
Bose-Hubbard model? 
Recent numerical studies beyond the Bose-Hubbard model (including higher-lying states in
the individual wells) \cite{Sakmann:2009wy,TrujilloMartinez:2009gh} show that ST is only present
as long as the system remains coherent.
Hence, there is evidence that our results do reflect a fundamental relation between ST and coherence of Bosonic quantum systems. 

In larger optical lattices, the self-trapped states of the dimer 
correspond to `discrete breathers' or `intrinsic localized modes' 
\cite{Campbell:2004vz,Flach:1998ul,Flach:2008ud}). 
These correspond to classical trajectories which are practically
embedded on a two-dimensional torus in the high dimensional
phase space and are thus (quasi-)periodic in time 
\cite{Flach:1993vb,Flach:1998ul,Flach:2008ud,Hennig:2010gy}.
In other words, discrete breathers involve localization in phase space.
Moreover, discrete breathers become attractive fixed points in presence
of dissipation and have been found in a variety of physical and biochemical systems 
\cite{Flach:2008ud,Campbell:2004vz}. Just as delocalization
in classical phase space leads to decoherence \cite{Mirbach:1995ul,Trimborn:2009up},
we expect that discrete breathers are candidates to support
long-lived coherent and possibly entangled many-body states even in 
complex dissipative systems, such as biomolecular systems.

\begin{acknowledgments}
We thank Ted Pudlik for comments. HH acknowledges financial support by the Deutsche Forschungsgemeinschaft
(DFG, grant no.~HE 6312/1-1). DKC thanks Boston University for partial support of this research and the Kavli Institute for Theoretical Physics for its hospitality during the completion of this work. 
\end{acknowledgments}

\bibliographystyle{apsrev}
\bibliography{}

\end{document}